# Visible light carrier generation in co-doped epitaxial titanate films

Ryan B. Comes[1], Sergey Y. Smolin[2], Tiffany C. Kaspar[1], Ran Gao[3], Brent A. Apgar[3,4], Lane W. Martin[3,5], Mark E. Bowden[6], Jason B. Baxter[2], and Scott A. Chambers[1*]

1Fundamental and Computational Sciences Directorate, Pacific Northwest National Lab, Richland, WA, USA
2Department of Chemical and Biological Engineering, Drexel University, Philadelphia, PA USA
3Department of Materials Science and Engineering, University of California-Berkeley, Berkeley, CA, USA
4Department of Materials Science and Engineering, University of Illinois at Urbana-Champaign, Champaign, IL USA
5Materials Science Division, Lawrence Berkeley National Laboratory, Berkeley, CA USA
6Environmental and Molecular Sciences Laboratory, Pacific Northwest National Lab, Richland, WA, USA

## Abstract

Perovskite titanates such as $SrTiO_3$ (STO) exhibit a wide range of important functional properties, including high electron mobility, ferroelectricity, and excellent photocatalytic performance. The wide optical band gap of titanates limits their use in these applications, however, making them ill-suited for integration into solar energy harvesting technologies. Our recent work has shown that by doping STO with equal concentrations of La and Cr we can enhance visible light absorption in epitaxial thin films while avoiding any compensating defects. In this work, we explore the optical properties of photoexcited carriers in these films. Using spectroscopic ellipsometry, we show that the $Cr^{3+}$ dopants, which produce electronic states immediately above the top of the O 2p valence band in STO reduce the direct band gap of the material from 3.75 eV to between 2.4 and 2.7 eV depending on doping levels. Transient reflectance spectroscopy measurements are in agreement with the observations from ellipsometry and confirm that optically generated carriers are present for longer than 2 ns. Finally, through photoelectrochemical methylene blue degradation measurements, we show that these co-doped films exhibit enhanced visible light photocatalysis when compared to pure STO.

Titanates perovskite such as $SrTiO_3$ (STO) and $BaTiO_3$ offer a variety of properties that are of interest for future electronic and energy applications. For example, electron-doped $SrTiO_3$ epitaxial thin films have been shown to exhibit the highest carrier mobility of any perovskite oxide.[1] STO and $BaTiO_3$ also exhibit enhanced ferroelectric behavior when grown with epitaxial strain on various substrates.[2,3] The presence of ferroelectricity, along with the good electron mobility, makes these materials good candidates for ferroelectric photovoltaics.[4] However, the wide band gap of both materials (3.25 eV indirect, 3.75 eV direct)[5] limits their applications in this regard. This large band gap also limits their use in solar hydrolysis where they would otherwise be ideal catalysts, given that the Ti $3d$ conduction band in STO is well aligned with the half-cell reaction energy to split $H_2O$ to $H_2$.[6]

Given the existing limitations of perovskite titanate materials, research on ferroelectric photovoltaics has focused on materials with band gaps in the visible light regime. $BiFeO_3$ thin films, with a band gap of 2.67 eV, have been shown to exhibit photoconductivity and photovoltaic behavior under visible light.[7,8] However, $BiFeO_3$ is a material known to produce leakage currents due to oxygen vacancies and other defects that are common.[9] Alternative routes to achieve visible light absorption in ferroelectrics have focused on substitution of a transition metal cation with a partially filled $3d$ orbital in place of the $d^0$ transition metal B site cation, such as Ti or Nb. In these systems, the $3d$ electrons lie at higher energies than the O $2p$ valence band maximum (VBM), raising the VBM of the material. Such an approach has been applied in Aurivillius $Bi_4Ti_3O_{12}$ films, with $La^{3+}$ dopants substituting for $Bi^{3+}$ and $Co^{2+}$ dopants substituting for $Ti^{4+}$ while also creating an oxygen vacancy to maintain charge neutrality.[10] This produced optical absorption at energies as low as 2.65 eV, similar to that of BFO. Recent work to dopebulk ferroelectric $KNbO_3$ has also been successful in reducing the band gap to 1.18 eV through stoichiometric alloying with $BaNi_{0.5}Nb_{0.5}O_{3-\delta}$ where $\delta$ would be equal to 0.5 in the case of the expected $Ni^{2+}$ and $Nb^{5+}$ oxidation states.[11] In each of these cases, however, an oxygen vacancy is created to maintain charge neutrality in the material, which could reduce mobility through defect scattering.

Work exploring photocatalytic applications of titanates offers a viable route to prevent the defects produced when a transition metal cation with a different oxidation state is doped into the material. Through the addition of a compensating A-site dopant, it is possible to maintain charge neutrality in the material without creating an oxygen vacancy. This has been demonstrated in $Cr^{3+}$-doped STO, where the substitution of equal concentrations of $La^{3+}$ donors onto the $Sr^{2+}$ site produces stoichiometric $Sr_{1-x}La_xTi_{1-x}Cr_xO_3$ (SLTCO).[12,13] In powder samples, visible light absorption and enhanced solar hydrolysis were observed in SLTCO, with a reported optical band gap of 2.12 eV.[14] We recently demonstrated that epitaxial SLTCO thin films could be prepared using oxide molecular beam epitaxy and showed that these films have ideal oxygen stoichiometry, thereby exhibiting visible light absorption at energies as low as 2.3 eV.[15] This approach offers a viable route to synthesize ferroelectric titanate photovoltaic thin films free of compensating defects. Here we characterize the optical generation of electron-hole pairs in SLTCO films and show that these materials exhibit enhanced photocatalytic behavior.

Epitaxial SLTCO films were grown on $(LaAlO_3)_{0.3}(Sr_2AlTaO_6)_{0.7}$ (LSAT) substrates using oxide molecular beam epitaxy via a shuttered growth technique described in the supplementary information,[16] with film stoichiometries controlled to within 1-3 atomic percent for Sr and Ti, and La:Cr ratio to ~5%. Homogeneous SLTCO films with doping levels, $x$, of 0.03 (20 nm) and 0.10 (25 nm) were grown at 700°C in $3\times10^{-6}$ Torr molecular $O_2$ at a rate of 2.7 Å/min, and were capped with 2 unit cells (u.c.) of STO to prevent over-oxidation of Cr during the cooling process. STO control films were grown using an electron cyclotron resonance microwave oxygen plasma source to ensure full oxidation, though this is not absolutely necessary for the growth conditions.[18] An additional 20 nm thick 10% SLTCO film was fabricated for photoelectrochemical (PEC) experiments with a 5 nm thick 3% La-doped STO bottom electrode and a 5 nm thick STO capping layer. A 20 nm thick STO film with the same bottom electrode was also fabricated for comparison. Representative x-ray diffraction scans are shown in the supplementary information.[16] All films were coherently strained to the substrate with out-of-plane lattice parameters consistent with stoichiometric STO films on LSAT.[19]

Films were characterized via variable angle spectroscopic ellipsometry to determine the optical properties as a function of photon energy. Ultrafast pump-probe transient reflectance (TR) spectroscopy was performed to measure photoexcited carrier dynamics in the films using a technique described previously.[20] PEC experiments were performed using a setup and measurement process that have been described previously.[21,22] Photocatalytic activities of the samples were examined by measuring the degradation rate of methylene blue solution, which is an effective method to evaluate the visible light absorption ability of the thin film material. Typically, the STO control sample and 10% doped SLTCO with 5nm STO protection layer were measured under an AM1.5G spectrum light filtered by a λ=444nm long-pass filter back-to-back for 12 hours. The filter cuts off light with wavelength shorter than 444nm (>~ 2.8 eV), thus eliminating effects of the primary O 2$p$ to Ti 3$d$ absorption in STO.

Fitted values for the extinction coefficient, $k$, and index of refraction, $n$, from the ellipsometry measurements are shown in Figure 1(a), while fits to the Tauc models for the direct allowed and direct forbidden band gap[23] are shown in Figure 1(b-c). The extinction coefficient was converted to the absorption coefficient, α, as $\alpha = 4\pi k / \lambda$, where λ is wavelength of the incident light. For the direct allowed band gap, the primary optical transition for the doped and undoped films is expected to be an excitation from the O 2$p$ valence band states to the Ti 3$d$ conduction band. The calculated gap for this transition is determined by extrapolating the linear portion of the graph of $(\alpha h v)^2$ as a function of incident photon energy, $hv$, and measuring the intercept with the energy axis. For the direct forbidden transition from the Cr 3$d$ dopant valence electrons to the Ti 3$d$ conduction band, the same method is used but the intercept of $(\alpha h v)^{3/2}$ is calculated. We find that the gap for the O 2$p$→ Ti 3$d$ transition increases with La and Cr doping levels, from 3.83(2) eV in STO to 3.89(2) eV and 4.06(2) eV for 3% and 10% doped films. Meanwhile, the direct gaps for the Cr 3$d$ to Ti 3$d$ transition are 2.47(2) eV and 2.66(2) eV for 3% and 10% doping, respectively. There was no evidence of indirect transitions at lower energies in the data, but the relatively weak absorption of an indirect forbidden transition makes it difficult to exclude the possibility of a lower indirect band gap. The values of $k$ and $\alpha$ at energies below the band gap of STO increase with increased doping levels, indicating an increased Cr 3*d*-derived density of states above the O 2*p* band.

The increase in the direct O 2*p* → Ti 3*d* gap with doping can be attributed to a reduction in the density of states at the bottom of the Ti 3*d*-derived conduction band with increasing doping levels. This behavior has been predicted theoretically using a density functional theory model of the SLTCO system for various doping levels, which showed that the density of states at the bottom of the conduction band was reduced.[24] The measured increase in the gap is qualitatively similar to the Burstein-Moss effect that occurs due to band filling of low-lying conduction band states.[25] However, we believe that a change in the conduction band density of states better explains the observations for two reasons: 1) the films are highly insulating; and 2) there is no evidence of a low-energy Drude peak in the ellipsometry data, suggesting the absence of free carriers in the conduction band. The magnitude of the change in the Cr 3*d* → Ti 3*d* direct gap energy from the 3% doped film to the 10% doped film was nearly identical to the change in the gap for the O 2*p* → Ti 3*d* transition, suggesting that the raised conduction band minimum contributes to the increased gap for this transition as well. The Cr 3*d* band is expected to have minimal dispersion due to the localized nature of the dopants at low concentrations,[24] so it is reasonable to expect that the dopants will produce either a direct or nearly direct gap material.

To further examine the nature of the optical excitations in SLTCO, ultrafast pump-probe transient reflectance spectroscopy was performed to measure the lifetimes and spectral response of photoexcited carriers. A detailed overview of the experimental setup has been described previously.[20] A 4.0 eV pump pulse was used to photoexcite electrons into the conduction band, and a white light probe pulse with a spectrum of 1.8-3.8 eV was used to monitor populations of photoexcited carriers with time resolution of 50 femtoseconds and range of several nanoseconds. Every other pump pulse was blocked with an optical chopper so that the photoexcited properties could be measured relative to a non-photoexcited reference. The change in reflectance of the film upon photoexcitation is proportional to the change in the refractive index due to photoexcited carriers.[26]

A color map of the change in reflectivity as a function of probe energy and delay time is shown in Figure 2(a) for the 3% doped film. Similar figures for STO and the 10% doped film are shown in supplementary materials.[16] Cuts from these maps are shown in Figure 2(b) and 2(c) to facilitate comparison of spectral and kinetic responses of the three samples. The SLTCO films show distinct transient reflectance feature between 2.5 and 3.0 eV that is consistent with the ellipsometric data in Figure 1(c), supporting the onset of the Cr $3d \rightarrow$ Ti $3d$ transition in that range. The local minima are ~2.8 eV and ~3.0 eV for the 3% and 10% samples, respectively, corresponding to a slightly larger band gap for the Cr $3d \rightarrow$ Ti $3d$ transition in films with increased doping. Photoexcitation at 4 eV generates carriers in higher-lying states, which appear to cool to this defect band within hundreds of femtoseconds, as indicated by the time scale of the Figure 2(c) inset. As expected, the undoped STO film did not exhibit this distinct band. Instead, the STO film exhibits a broad reflectance transient that spans the width of our detection range. Similar features have been reported previously for STO[27–30] and attributed to a range of causes including defect bands,[28] intraband transitions[29], and hole polarons[30]. Features consistent with oxygen vacancies or free carriers that have been observed in other pump-probe measurements were not observed for either the STO sample or the doped films, in agreement with the ellipsometry data that showed no Drude response.[29]

Carrier dynamics were evaluated by tracking the transient reflectivity of each sample at the probe energy corresponding to its minimum, $\Delta R$, as shown in Figure 2(c) which plots $\Delta R$ normalized by its maximum value near t=0. In this regime, $|\Delta R/\Delta R_{max}|$ is equivalent to $n/n_i$, where $n$ and $n_i$ are the carrier density and initial carrier density. The decay kinetics show two main features: a strong decay in which the majority of carriers recombine within 10 ps, and a slower decay that extends into the nanosecond range. Roughly 10% of carriers are still present after 2 ns.

The kinetics were fit well using the Auger and Shockley-Read-Hall (SRH) recombination models:

$$\left|\frac{\Delta R}{\Delta R_{max}}\right| = A_1(1 + k_1 t)^{-\frac{1}{2}} + A_2 \exp\left(-\frac{t}{\tau_2}\right). \qquad (1)$$

The first term arises from Auger recombinationand captures the dynamics within the first few picoseconds when carrier densities are very high. The second term arises from first-order recombination processes such as Shockley-Read-Hall recombination, and captures the dynamics at longer times. The $A_i$ parameters represent the relative weights, $k_1$ is the rate constant of Auger recombination and is dependent upon the initial carrier density, and $\tau_2$ is time constant of SRH recombination. The fits are shown in Figure 2(c), fit parameters are given in Table I, and additional details regarding the model are provided in the supplementary materials.[16]

While other models might also be consistent with the data, this model is based on physically reasonable recombination mechanisms. According to previous work[31] and given the high excitation intensity of 1.06 mJ/cm$^2$-pulse, the fast decay in the first few ps is likely due to Auger recombination. Indeed, when monitoring kinetics within only this time window, the rate of decay was best represented by only the Auger recombination term, as shown in the supplementary materials.[16] Auger recombination lifetimes correlate inversely with carrier densityof photoexcited holes in the Cr 3d band. The 10%-doped film is likely to have a higher density of Cr 3d states than the 3%-doped film, which is consistent with the faster early-time kinetics observed in the more highly doped film.

The exponential decay had time constant, $\tau_2$, on the order of nanoseconds for both samples. The exponential decay model is consistent with SRH recombination. The more highly doped SLTCO sample showed a shorter SRH time constant, which could be due to higher concentrations of mid-gap defect states with increased doping. This exponential decay could also arise from surface recombination or other mechanisms. Regardless, the slow decay process enables a significant fraction of long-lived carriers that may be harnessed for photovoltaic or photocatalytic applications.

PEC measurements were also performed on two separate samples prepared with La-doped STO bottom electrodes. The normalized absorbance of the methylene blue dye was plotted with respect to the elapsed measuring time in Figure 3, with a 5-point boxcar average to reduce noise. The gray curve represents the spontaneous degradation of MB solution under the visible portion of the AM1.5G simulated solar light beamwith a glass coupon in place of the sample. By fitting the curve and subtracting the spontaneous degradation rate constant, we obtain the absolute rate constants of the STO and SLTCO samples. Both film samples showed positive photocatalytic activities relative to the glass control and that La, Crdoped STO exhibits greater photocatalytic performance compared to the un-doped STO when exposed to visible light. This is likely a result of the reduced band-gap in the doped samples, as has been seen in experiments performed on SLTCO powders previously.[13,14] After approximately 9 hours, the absorbance for the STO, SLTCO and glass slides begin to converge over the next ~3 hours until they are equal. The convergence of the film samples with the control may be explained by damage to the surface in the STO film and 5 nm STO capping layer for the SLTCO sample. Surface damage in the form of reduction in the STO cap or gradual etching and removal of the films would likely irreversibly reduce catalytic performance in the films. Repeated measurements of the SLTCO film showed that it did not exhibit photocatalytic behavior beyond that of the glass slide, supporting this hypothesis.

In summary, we have explored optical carrier generation in La, Cr co-doped $SrTiO_3$ epitaxial thin films using a variety of characterization techniques. Spectroscopic ellipsometry confirmed that the direct band gap of the doped films is reduced by the addition of stable $Cr^{3+}$ dopants, and that the reduction in the band gap varies inversely with doping levels. This is attributed to the removal of the lowest lying Ti conduction band states as more Cr ions are substituted for Ti ions on the B site, and is supported by measurements of the direct gap for the O $2p \rightarrow$ Ti $3d$ transition. Transient reflectivity measurements showed an absorption band in the energy range expected for Cr $3d \rightarrow$ Ti $3d$ transitions and agreed with the ellipsometry measurements. We find that increases in Cr doping concentration enhance the photoexcited carrier density based on the reduction in the Auger recombination lifetime with doping. Other carrier lifetimes were comparable between the doped films, suggesting long-lived photoexcited carriers due to the presence of stable $Cr^{3+}$ dopants. Photoelectrochemical yield measurements confirmed that the addition of the Cr dopants enhanced visible light photocatalysis in the films when compared to STO.

**Acknowledgements**

The authors would like to thank Prof. Steve May for helpful discussions and for providing single crystal substrates for transient reflectance calibration. RBC was supported by the Linus Pauling Distinguished Post-doctoral Fellowship at Pacific Northwest National Laboratory (PNNL LDRD PN13100/2581). SAC, TCK, and MEB were supported at PNNL by the U.S. Department of Energy, Office of Science, Division of Materials Sciences and Engineering under Award #10122. The PNNL work was performed in the Environmental Molecular Sciences Laboratory (EMSL), a national science user facility sponsored by the Department of Energy's Office of Biological and Environmental Research and located at Pacific Northwest National Laboratory. The Drexel work was supported by the National Science Foundation through award ECCS-1201957 using an ultrafast spectrometer that was acquired with funds from NSF MRI award DMR-0922929. RG acknowledges support from the Air Force Office of Scientific Research under grant FA9550-12-1-0471. BAA acknowledges support from the International Institute for Carbon-Neutral Energy Research (WPI-I2CNER), sponsored by the Japanese Ministryof Education, Culture, Sport, Science and Technology. LWM acknowledges support from the National Science Foundation under grant DMR-1124696.

**Table 1: Fitted parameters for kinetics using an Auger and SRH recombination model.**

| Sample | Auger | | SRH | |
|---|---|---|---|---|
| | $A_1$ | $k_1(\text{ps}^{-1})$ | $A_2$ | $\tau_2(\text{ps})$ |
| SLTCO 3% | 0.98±0.02 | 3.28±0.18 | 0.09±0.004 | 3284±787 |
| SLTCO 10% | 0.79±0.02 | 4.18±0.31 | 0.20±.004 | 964±66 |

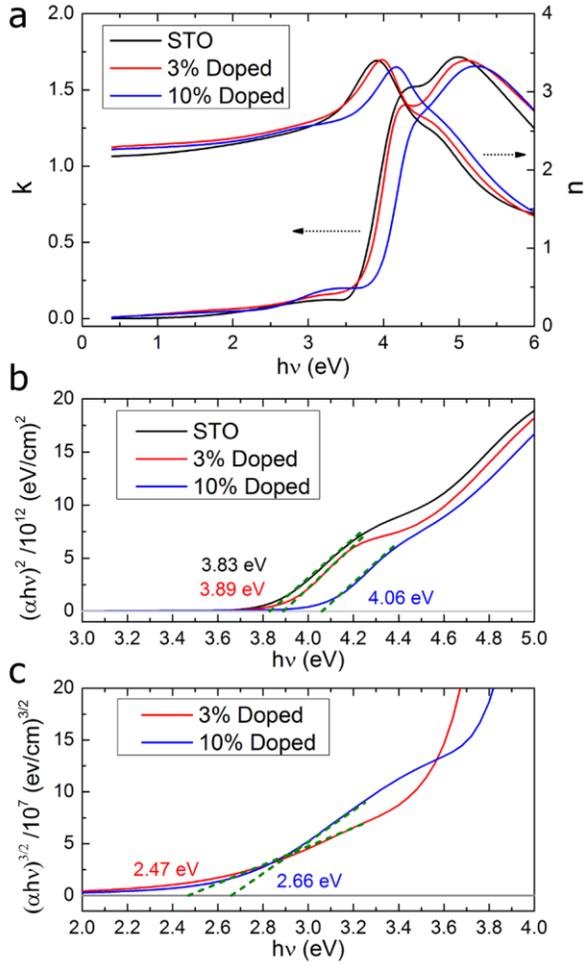

Figure 1: Ellipsometry data for SrTiO$_3$ and SLTCO. a) Extinction coefficient, $k$, and index of refraction, $n$; b) Fits measuring the direct band gap for O 2$p$→Ti 3$d$ transitions for all three films; c) Fits measuring the direct band gap for Cr 3$d$→Ti 3$d$ transitions for doped films.

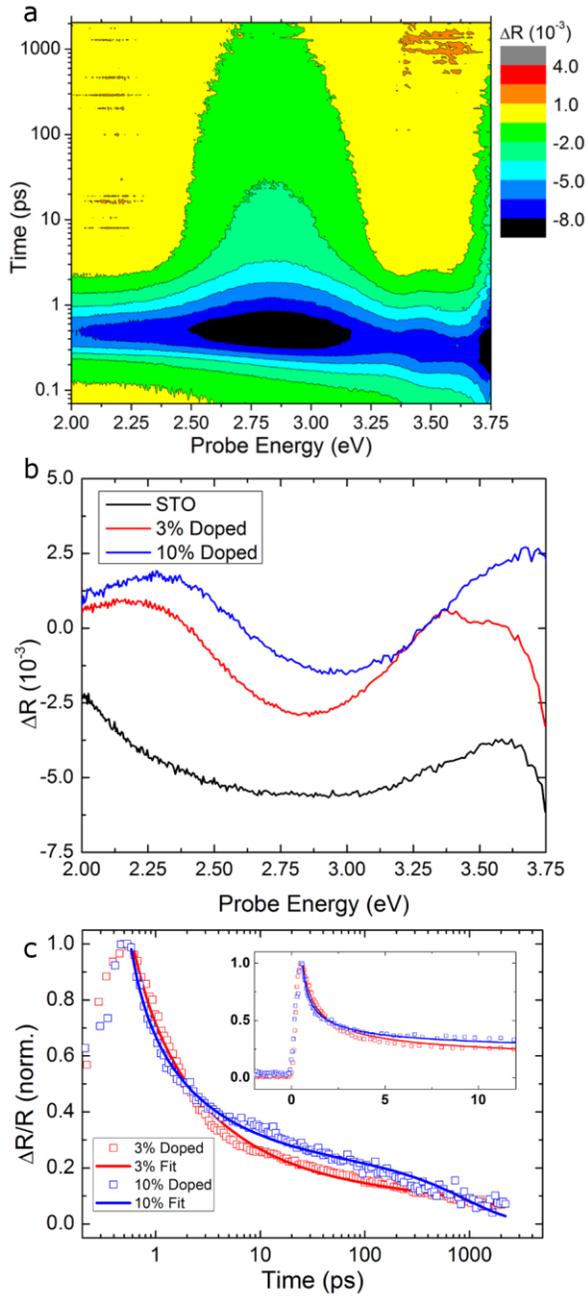

Figure 2: a) Map of reflectivity change with time and probe energy; b) Spectral response at ~5 ps; c) Normalized transient reflectivity at energy corresponding to the minimum ΔR and model fit. All films were pumped at 4 eV with 1.06±0.04 mJ/cm$^2$-pulse.

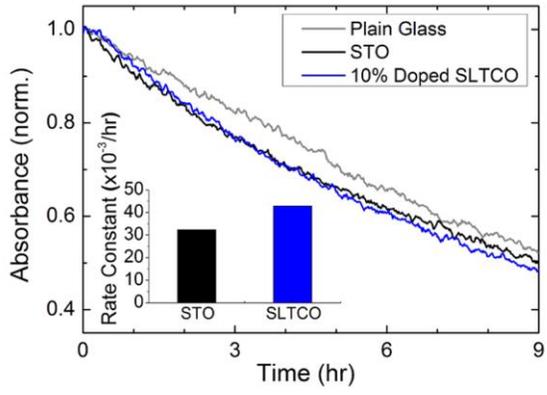

**Figure 3:** MB Absorbance versus time, and degradation rate constants, for films and reference glass sample .

# Visible light carrier generation in co-doped epitaxial titanate films

Ryan B. Comes[1], Sergey Y. Smolin[2], Tiffany C. Kaspar[1], Ran Gao[3], Brent A. Apgar[3,4], Lane W. Martin[3,5], Mark E. Bowden[6], Jason B. Baxter[2], and Scott A. Chambers[1*]

1 Fundamental and Computational Sciences Directorate, Pacific Northwest National Lab, Richland, WA, USA
2 Department of Chemical and Biological Engineering, Drexel University, Philadelphia, PA USA
3 Department of Materials Science and Engineering, University of California-Berkeley, Berkeley, CA, USA
4 Department of Materials Science and Engineering, University of Illinois at Urbana-Champaign, Champaign, IL USA
5 Materials Science Division, Lawrence Berkeley National Laboratory, Berkeley, CA USA
6 Environmental and Molecular Sciences Laboratory, Pacific Northwest National Lab, Richland, WA, USA

## Supplemental Online Information

**Film Growth**

Films were grown using oxide molecular beam epitaxy via a shuttered growth technique reported previously.[1] In summary, an initial $SrTiO_3$ (STO) homoepitaxial thin film is grown on an STO substrate via a shuttered growth technique from Sr and Ti elemental sources in effusion cells. The cell temperatures are coarsely adjusted using a quartz crystal microbalance to known calibration standards prior to growth. The homoepitaxial film is then grown at 700 °C in $3\times10^{-6}$ T of background oxygen pressure with atomic oxygen generated via an electron cyclotron resonance (ECR) plasma. The growth is monitored via in situ reflection high energy electron diffraction (RHEED) and the cell temperatures are adjusted until RHEED oscillations are constant throughout the growth. This technique has previously been reported to produce ideal Sr:Ti ratios to within 1-3%.[2] After the homoepitaxial calibration, the $(LaAlO_3)_{0.3}(Sr_2AlTaO_6)_{0.7}$ (LSAT) substrate is loaded into the chamber and a La, Cr co-doped STO film (SLTCO) is grown at the same temperature but without the ECR plasma due to the propensity of Cr ions to overoxidize in plasma.[3] Films grown using this technique have been shown to have ideal oxygen stoichiometry based on x-ray photoelectron spectroscopy measurements.[1] Representative out-of-plane x-ray diffraction and reciprocal space map data for the 3% doped SLTCO film are shown below in Figure S1. The out-of-plane lattice parameters for all films reported in this work were consistent with nominally stoichiometric STO coherently strained to an LSAT substrate (a = 3.87 Å).[4] Fits to the thickness oscillations about the (002) peak confirmed that the film thickness matched that calibrated from RHEED oscillations to within 0.5 nm.

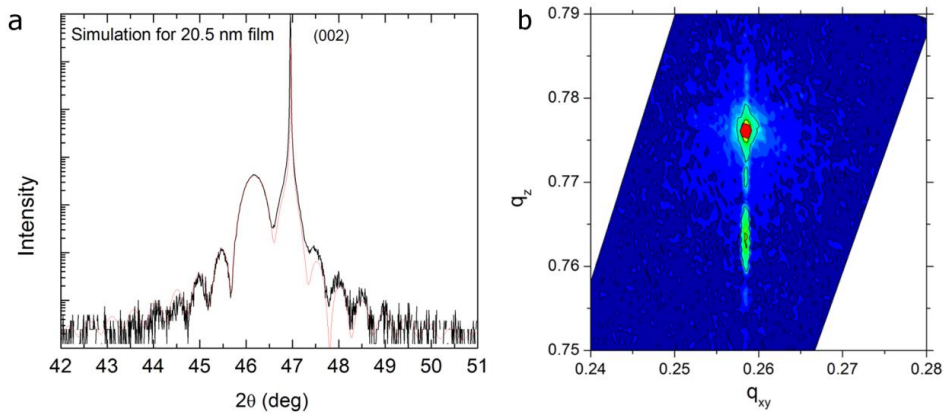

**Figure S1: a) Out-of-plane x-ray diffraction scan about (002) peak of film and substrate with model fit; b) Reciprocal space map about (103) peak of film and substrate.**

## Transient Reflectance

Transient refelectance (TR) measurements were performed on all three samples discussed in the main paper, but only the spectral color map for the 3% doped SLTCO sample was included in the text for space. Spectral color maps of the STO control sample and the 10%-doped SLTCO film are shown below in Figure S2 Figure S. For the STO film, a broad negative reflectance bleach is observed with no distinct energy bands in the region of interest. This bleach is consistent with what is seen in a pure STO single crystal substrate (see Figure S3), indicating that the STO film exhibits the expected behavior in the TR measurements. The 10% doped film exhibits a band of decreased reflectance between ~2.5 and ~3.25 eV, similar to what was seen in the 3% doped sample. No TR intensity was observed for the LSAT substrate, which has a wider optical band gap than the 4.0 eV pump laser beam, as can be seen in Figure S4.

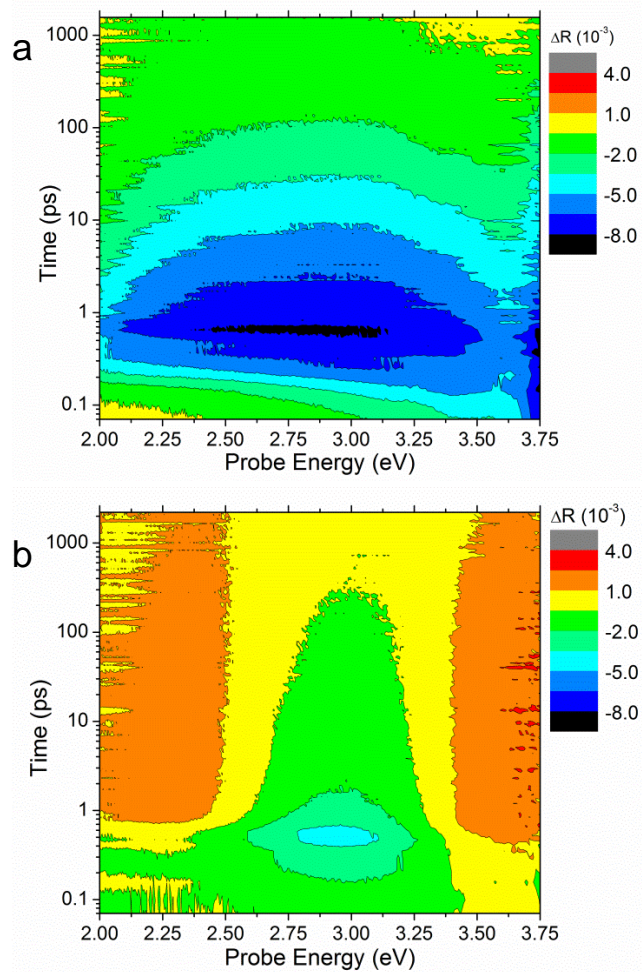

**Figure S2: Color maps of time and energy dependence of reflectivity for a) SrTiO3 and b) 10% doped SLTCO.**

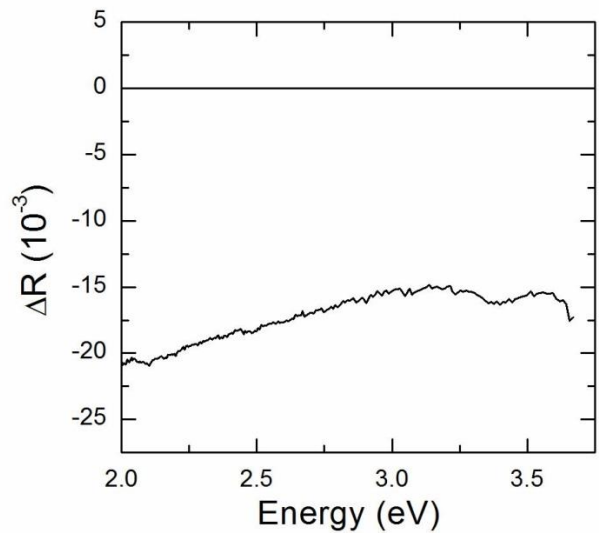

**Figure S3:** Reflectance versus energy for SrTiO$_3$ substrate showing the same broad bleach feature as in the STO film. The pump-probe time delay is 5 ps.

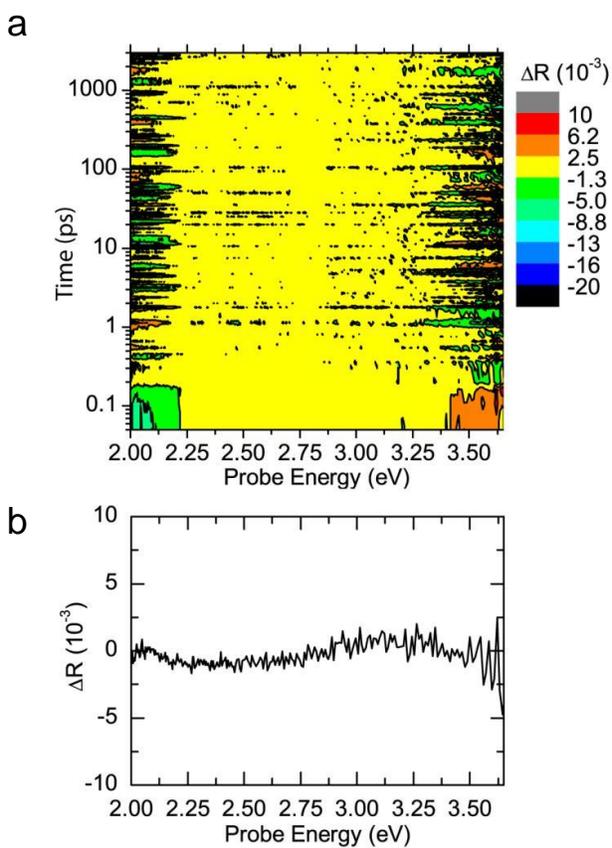

**Figure S4:** a) Reflectance transient color map and b) line profile for LSAT substrate showing no change in reflectance.

**Fitting**

Reflectance transients were fit with anAuger-SRH model:

$$\left|\frac{\Delta R}{\Delta R_{max}}\right| = A_1(1 + k_1 t)^{-\frac{1}{2}} + A_2 \exp\left(-\frac{t}{\tau_2}\right)$$

This four-parameter model fit the data well and is derived from a physical understanding of the material system. The reflectance transient was monitored at its local minima and normalized by its maximum magnitude.

The rate of in Auger recombination is given as $\frac{dn}{dt} = -Cn^3$, whose solution is $\left(\frac{n}{n_i}\right) = \frac{1}{(1+n_i^2 Ct)^{\frac{1}{2}}}$. Where C isa fit parameter and $n_i$ is the initial photoexcited carrier density. Analogously, the rate of carrier decay in SRH recombination is given as $\frac{dn}{dt} = -An$, whose solution is $\frac{n}{n_i} = \exp\left(-\frac{t}{\tau}\right)$, where $\tau = A^{-1}$. Assuming that the two recombination mechanisms occur on different time scales, a linear combination of them is justified and provides clear identification of each mechanism. Moreover, because $\left(\frac{n}{n_i}\right) = \left(\left|\frac{\Delta R}{\Delta R_{max}}\right|\right)$, it is possible to fit the reflectance data to the recombination model mentioned above. In fitting the Auger model $k = n_i^2 C$ was used to simplify the fitting procedure.

**Auger recombination rate dependence in the first few ps**

As mentioned above, the rate of carrier decay in Auger recombination is given as $\frac{dn}{dt} = Cn^3$. Solving this equation and linearizing gives $\left(\frac{n_i}{n}\right)^2 = A + 2n_0^2 Ct$. Because $\left(\frac{n}{n_i}\right) = \left(\frac{\Delta R}{\Delta R_{max}}\right)$, a plot of $\left(\frac{\Delta R_{max}}{\Delta R}\right)^2$ vs $t$ will be linear if the data are consistent with this model, as shown in Figure S5(c). Also, when compared to other linearized expressions for Shockley-Read-Hall (SRH), and radiative recombination, as shown in Figure S5 (a)-(b), Auger recombination results in the most linear fit in the first few ps, suggesting that Auger recombination dominates the decay in the first few ps.

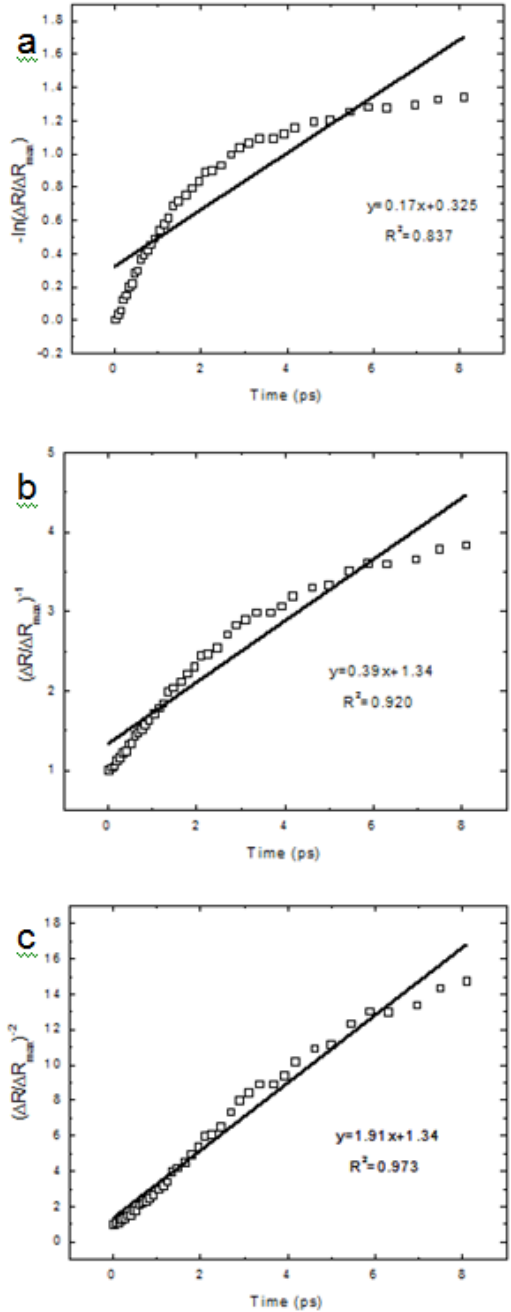

**Figure S5: linearized kinetic data for (a) SRH, (b) radiative recombination, and (c) Auger recombination. Data from 3% SLTCO sample.**